				\documentstyle[12pt,aaspp4]{article}
				
\begin{document}
				\title{
$ROSAT$ Temperatures and Abundances for a Complete Sample of 
Elliptical Galaxies
				}
				\author{
David S. Davis \& Raymond E. White III     
				}

                                 \affil{
Department of Physics \& Astronomy, University of Alabama, Tuscaloosa, 
AL 35487-0324   
				}

                                \begin{abstract}
We determine X-ray temperatures and abundances for elliptical galaxies
drawn from a complete, optically selected sample.  
The optical magnitude-limited sample consists of 43 galaxies, complete to a  
(corrected) $B$ magnitude of $m_B^o=11.36$.
Of these, 30 have enough X-ray spectral counts to allow temperature
determinations.
We find that the temperatures of the X-ray emitting gas in these 
ellipticals are correlated with the central stellar velocity dispersions:
$T\propto \sigma^{1.45}$; this is a shallower trend than the simple 
thermal relation $T\propto\sigma^2$.
The diffuse gas is substantially hotter than the kinetic 
temperature of the luminous stars, since $kT>\mu m_p\sigma^2$ for all 
galaxies with measurable temperatures.
This strongly indicates that dark matter halos are characterized by 
velocity dispersions which exceed those of the luminous stars.
We see no evidence of emission from X-ray binaries becoming progressively
more dominant in lower luminosity ellipticals.
We find that so-called ``supersoft" sources adhere to the observed $kT-\sigma$ 
relation, so they are no softer than expected for their velocity dispersions.
We also find that ISM temperatures and abundances are correlated, 
with the gas in hotter systems being more enriched than in cooler galaxies.  
However, no correlation is found between gaseous abundances 
and stellar abundances, as inferred from $Mg_2$ indices.
				\end{abstract}

				\keywords{
galaxies: abundances --- galaxies: elliptical --- 
galaxies: ISM --- X-ray: galaxies 
				}
				\clearpage

				\section{
Introduction
				}
Elliptical galaxies have been known as X-ray sources since
several were detected in the Virgo cluster by the
$Einstein$ $Observatory$ (Forman et al. 1979).
Long and Van Speybroeck (1983) compiled many more $Einstein$
detections of ellipticals and Nulsen, Stewart and Fabian (1984) 
showed that diffuse gas is their dominate source of X-ray emission. 
Analysis of $Einstein$ $Observatory$ IPC spectra shows that the emission
from luminous ellipticals can be 
fit by thermal plasma models with fairly low temperatures, 0.5 -- 
2.0 keV (Forman, Jones, \& Tucker 1985; Trinchieri, Fabbiano \& Canizares 1986).
More recent $Ginga$, $ROSAT$ and $ASCA$ observations of individual galaxies 
confirm these characteristic temperatures (cf. Awaki et al.1991; 
Forman et al. 1993; Fabbiano, Kim \& Trinchieri 1994; Awaki et al.1994; 
Matsushita et al.1994).
The emission from lower luminosity ellipticals is apparently more complex, 
with evidence for both harder and ``supersoft" emission according to
Fabbiano, Kim \& Trinchieri (1992); these authors attribute 
the hard component to stellar X-ray binaries and suggest the supersoft emission 
may be from M stars, ultrasoft X-ray binaries and/or relatively cool ISM.

To date, there has been no study of the X-ray properties of a $complete$ 
sample of normal elliptical galaxies.  The IPC detector on the $Einstein$ 
$Observatory$ was used to observe $\sim200$ normal galaxies, 
including many ellipticals (see the review by Fabbiano 1989), but these data
allow only small ($N\approx14$) complete samples of ellipticals to be 
constructed.
Furthermore, the limited energy resolution of the IPC prevented accurate
temperature determinations.
We have therefore turned to the longer exposures of $ROSAT$ pointed 
observations to construct the largest possible complete sample of accurate 
X-ray luminosities and temperatures for elliptical galaxies.
In this letter we 
present new gas temperature and abundance determinations from
$ROSAT$ PSPC spectra of a subset of this complete sample
and explore their relation to optical and dynamical 
properties of the host galaxies.  We will present detailed analyses
of their X-ray luminosities in a forthcoming paper (Davis \& White 1996).

				\section{
Sample Selection                
				}
We used the $HEASARC$ $BROWSE$ software to select our optical magnitude-limited 
sample from the $Third$ $Revised$ $Catalog$ $of$ $Galaxies$ (de Vaucouleurs 
et al. 1991; hereafter {\sl RC3}), using its corrected $B$ magnitudes.
We limited our sample to elliptical galaxies (de Vaucouleurs T-types -4, -5 
\& -6).
We then used $BROWSE$ to cross-correlate this sample with the $ROSAT$ and 
$Einstein$ pointed observations.  Neglecting galaxies in the zone-of avoidance 
(taken here to be $|b|<20^\circ$), we found that 38 of the 43 optically 
brightest ellipticals have $ROSAT$ PSPC data, and 4 of the remaining 5 have 
$Einstein$ IPC data (the remaining galaxy, NGC 7507, will soon have $ROSAT$ HRI 
data available, but this has no spectral information);  since $ROSAT$ PSPC data 
have better spectral resolution than $Einstein$ IPC data, they are preferable 
when there is a choice.  
The sample of 43 optically-selected ellipticals is complete to a limiting 
(corrected) blue magnitude of $m_B^0=11.36$ and includes  
four Local Group dwarfs.
The X-ray counterpart to our optical sample is thus 98\% complete, with X-ray
data for 42 of the 43 ellipticals.  
If we press to fainter optical magnitudes, the marginal completeness in the 
X-ray data deteriorates at the 50\% level for the next 10 or so optically 
brightest galaxies.
In this letter we present a subset of 30 galaxies from this complete sample 
which have sufficient spectral counts to accurately determine their X-ray 
temperatures; 27 of these galaxies also have well-constrained abundances.

				\section{
X-Ray Data                
				}
In order to tie the X-ray properties of ellipticals to their optical properties
in a uniform manner, we let the optical photometry dictate the spatial size of 
the X-ray spectral extraction regions.
For each galaxy in our optically selected sample, we extract X-ray spectra from 
within $6r_e$, where $r_e$ is the galaxy's optical effective radius 
(derived from the {\sl RC3}).
Due to X-ray scattering in the $ROSAT$ mirror assembly, the minimum 
suggested extraction radius for point sources is $1.5^\prime$, which is the
99.5\% encircled energy radius for a 1 keV spectrum.  
Our minimum metric extraction radius must be $3r_e$ to meet this requirement 
with the current sample. 
The purpose of our larger metric extraction radius of $6r_e$ is to provide
more global X-ray properties, for better comparison with global 
optical quantities such as magnitudes 
($6r_e$ encloses 91\% of a de Vaucouleurs $r^{1/4}$ profile).

For the PSPC data, source spectra were selected from observation times when the 
{\it Master Veto Rate} ($MVR$) was within recommended bounds: $20<MVR<170$.  
The background for each spectrum was obtained from an annulus around 
the galaxy extending from $6r_e$ to $12r_e$ from its center.
All detectable point sources were excluded from both source and background 
spectra;
excluded points in the source regions tended to be in the outskirts so they 
are unlikely to be X-ray binaries. Source and background spectra were also 
corrected for vignetting effects and for the residual 
particle background (in the manner prescribed by Snowden et al. 1994),
which is proportional to the $MVR$. 
To ensure correct Poisson errors were used throughout, we modified the 
appropriate data header as outlined in the $PROS$ QPSPEC documentation.
Finally, the extracted spectra were rebinned so that each channel has a 
minimum of 25 counts, and systematic errors of 1\% were included. 

				\section{
Analysis
				}
We used {\sl XSPEC~9.0} software (Schafer et al. 1990) to fit various models to 
the extracted spectra.
An isothermal Raymond-Smith plasma model was our benchmark model,
with the temperature and abundance allowed to vary in the fits.
We also included a variable absorption component due to the column density of 
Galactic hydrogen in the line-of-sight.
The redshift of the model spectrum was fixed to that corresponding to the 
heliocentric velocity of the galaxy given in the {\sl RC3}. 
The extracted spectra were fit between $\sim$0.2 and 2.0 keV, the exact energy 
boundaries being set by the channel grouping. 
Once a minimum in $\chi^2$ was found, the 90\% confidence
errors were determined for the free parameters. 
Most of the galaxies have sufficient
counts so that the temperature, abundance and absorption could all be fit. 
For a few galaxies, however, the number of free parameters had to be reduced to 
allow reasonable errors to be determined. 
For such galaxies we first fixed the abundance to 0.2 
solar; when necessary, we also fixed the absorption to the Galactic value 
(Stark et al. 1992). 
The Raymond-Smith thermal model provided formally excellent fits to all 
but two galaxies, with $\chi^2$ per degree of freedom $\nu$ of 
$\chi^2_\nu\approx1$. 
The two exceptions (NGC~4486 \& IC~1459) are known to have AGN, 
so we fit composite models consisting of a Raymond-Smith isothermal
plus a power-law spectral component; these composite spectra achieved 
$\chi^2_\nu\approx1$.
The detailed spectral fitting results are shown in Table~1, where the
indicated errors are 90\% confidence limits.

				\subsection{
$kT$--$\sigma$ Correlation
				}

For the 30 galaxies which have enough counts to allow accurate 
temperature determinations, Fig.~1 shows a logarithmic plot of their 
temperatures $kT$ against $\sigma$, the central 
line-of-sight stellar velocity dispersions within the galaxies (from Faber 
et al. 1989 or Dressler, Faber \& Burstein 1991).  
Characteristics of the best fit linear regression to these data are
shown in the first line of Table~2.  We found that three galaxies (M32, 
NGC 1052 and NGC 2768) have temperatures of $\sim3-4$ keV, which are
much greater than those of other galaxies with similar velocity dispersions. 
These galaxies are likely to have their X-ray emission dominated by X-ray 
binaries or AGN (cf. Eskridge, White \& Davis 1996; 
Ho, Filipenko \& Sargent 1995), rather than by the cooler diffuse gas 
expected from the accumulation of stellar ejecta.
Neglecting these three galaxies, 
the best fit linear regression yields $kT \propto \sigma^{1.45\pm 0.20}$ and
the temperature dispersion is only 0.10 dex (see line 2 of Table~2);
the regression is shown as a solid line in Figure~1. 
This correlation is shallower than what would obtain if the gas 
and stars had the same kinetic temperature: $kT=\mu m_p\sigma^2$; this latter
relation is also shown in Fig.~1 as a dashed line.
The bulk of our temperature measurements indicate that warm diffuse gas
is being detected, even down to low luminosities and velocity dispersions
where some investigators expect such gas to be swept by winds (e.g. David, 
Forman \& Jones 1991; Ciotti et al. 1991).
Fig.~1 also shows that the global gas temperatures are hotter than the kinetic 
temperatures of luminous stars in the galaxies, since $kT>\mu m_p \sigma^2$
in all the galaxies.

				\subsection{
Temperature--Abundance Correlation
				}
For those 27 galaxies with enough counts for elemental abundances to be 
constrained, we find that the measured abundances (driven by $\sim$1 keV Fe L 
emission in $ROSAT$ PSPC spectra) are correlated with the gas temperatures: 
hotter gas tends to have higher abundances (see Fig.~2). 
Line 3 of Table~2 shows the regression results for all the data.
When we exclude M32 and NGC 1052 from the regression (for being hot point
sources, rather than diffuse gas), we find the abundance $A \propto T^{2.44\pm
0.69}$ (see line 4 of Table~2), with a dispersion of 0.48 dex in $A$. 
This latter correlation is shown as a dotted line in Fig.~2 (where
M32 and NGC 1052 are also indicated).
We have checked our abundances against the several published $ASCA$ values
and find that in most cases they agree within the 90\% confidence 
level and in all cases they agree within the 99\% confidence level.  
Our PSPC temperatures are also consistent with the $ASCA$ results. 
We find no correlation between our measured abundances and the stellar 
abundances of the host galaxies, as inferred from $Mg_2$ indices (7S)
(see lines 5-6 of Table~2); 
this is consistent with the $ASCA$ results of Loewenstein et al. (1994) and 
Mushotzky et al. (1994). 

				\section{
Summary and Discussion
				}

In this letter we have shown that gas temperatures are correlated with stellar 
velocity dispersions, $T \propto \sigma^{1.45}$, and are everywhere hotter than 
the stellar kinetic temperatures ($kT > \mu m_p \sigma^2$).
There are several possible mechanisms for the gas acquiring more specific 
energy than the stars.
One possibility is that supernovae have heated the gas, but this is inconsistent
with the generally low metal abundances being found in such galaxies 
(Matsushita et al. 1994; Loewenstein et al. 1994; and \S4.2 of this paper).  
A more likely possibility is that the luminous parts of the galaxies are 
embedded in dark matter halos which are dynamically hotter than the stars
(Fabian et al. 1986).  
As stellar ejecta accumulates and flows inward in cooling flows,
movement through the gravitational potential heats the gas.
If dark matter halos are characterized by velocity dispersions that are 
$\ga50\%$ greater than those of the luminous stars, then gravitational
heating can raise the gas temperatures to the observed values.

In the largest study of X-ray spectral properties of galaxies to date,
Kim, Fabbiano \& Trinchieri (1992, hereafter KFT) found that the X-ray 
emission from elliptical galaxies is more complicated 
than the simple thermal plasma found by Forman et al. (1985). 
KFT found for galaxies with high X-ray to optical flux ratios ($L_X/L_B$)
that their spectra are indeed well described by Raymond-Smith thermal 
plasma models with temperatures of $\sim1$ keV and solar abundances. 
But for galaxies with intermediate values of $L_X/L_B$, KFT found spectral
hardening. At even lower values of $L_X/L_B$, the spectra require a 
a very cool ``supersoft" 0.2 keV component in addition to the hard 
($\sim$5 keV) component. 
Although a few galaxies in our sample have unusually hard emission, we find 
that the correlation of temperature with velocity dispersion persists down to 
$kT\approx0.2$ keV, $L_B\approx4.7\times10^9h^{-2}$ $L_\odot$, 
$\sigma\approx120$ km s$^{-1}$. 
Our sample includes some of the galaxies deemed to have ``supersoft" spectra
by KFT, but we find that these sources are simply on the cool end of a 
continuous distribution, with temperatures consistent with expectations  
from ISM in galaxies with low velocity dispersions 
(hence, shallower gravitational potentials).

While a few of the galaxies in our sample exhibit hard emission (due to X-ray 
binaries, AGN, or both) we find no general hardening of the X-ray emission 
as we go to lower luminosities and velocity dispersions, as might be expected 
if low-luminosity ellipticals are blowing winds. 
KFT find harder spectra for a subset of their sample (their ``Group 2"), which
they attribute to X-ray binaries, but our PSPC data for 4 of 8 galaxies in 
this subsample (after removing NGC~1052, a Seyfert 2, which may have biased
the results of KFT) show no evidence for 
having harder spectra attributable to X-ray binaries. 
Thus, we detect gaseous X-ray emission without contamination by X-ray binaries 
down to $L_B\approx4.7\times10^9h^{-2}$ $L_\odot$, 
$L_X\approx10^{39}h^{-2}$ erg s$^{-1}$ and $\sigma\approx120$ km s$^{-1}$. 
This is contrary to the predictions of some theoretical models for the
evolution of hot gas in ellipticals, in which the threshold for the onset
of supernovae-driven winds occurs at significantly higher optical luminosities
(Ciotti et al. 1991; David et al. 1991).  Such winds drive out the gas, leaving
only X-ray binaries as significant sources of X-ray emission.
M32 may be an example of such a wind-blowing galaxy, since it exhibits hard
X-ray emission and is ten times less luminous (both optically and in X-rays)
than the next less luminous galaxy in our sample. 

Finally, we find that abundances are correlated with global temperatures:
$A \propto T^{2.4}$. 
This may reflect an underlying trend in observed stellar abundances (more 
luminous galaxies are more metal rich), but we find no detailed correlation 
between our gaseous abundances and the stellar metallicity index $Mg_2$.
The $Mg_2$ index is measured within the inner few kpc of galaxies, so it may 
not be representative of the galaxy out to our extraction radius of $6 r_e$. 
For example, Davies, Sadler \& Peletier (1993) find that the $Mg_2$ index
drops by $\sim$25\% from the center to a radius of $\sim1 r_e$.
Analysis of $ASCA$ observations of the elliptical NGC 4636 shows there is a 
gaseous abundance gradient evident from spatially resolved X-ray spectroscopy,
a gradient which persists to at least $5 r_e$ (Mushotzky et al. 1994).
Our global, emission-weighted measures of abundances obscure such structure.
The lack of correlation between the stellar and gaseous abundances 
may alternatively indicate that more luminous galaxies have larger supernova 
rates than less luminous galaxies, but not so large as to unbind the gas. 

In a subsequent, more detailed analysis of this complete sample of ellipticals, 
we will present their X-ray luminosities, their luminosity function, discuss 
the Local Group dwarf galaxies, and assess additional correlations and 
implications.

\acknowledgments
This research made use of the HEASARC, NED, and SkyView databases. We would
like to thank Keith Arnaud for his help with the IPC data. 
This work was partially supported by the NSF and the State of Alabama through
EPSCoR grant EHR-9108761.  REW also acknowledges partial support from NASA 
grants NAG 5-1718 and NAG 5-1973.

%				\clearpage

				\clearpage

\begin{deluxetable}{lcccccc}
\small
\tablewidth{0pt}
\tablecaption{Spectral Fits\tablenotemark{a}}
\tablehead{
\colhead{Name} &  
\colhead{$kT$}  & 
\colhead{$Abundance$} &  
\colhead{$N_H$\tablenotemark{b}}  &  
\colhead{$N_H$\tablenotemark{b}}  &  
\colhead{$\chi^2/\nu$}  \nl
\colhead{} &  
\colhead{(keV)}  & 
\colhead{(solar)} &  
\colhead{fitted}  &  
\colhead{Galactic}  &  
\colhead{} 
}
\startdata
N  221 & 3.94$^{+1.73}_{-1.06}$ & 0.44$^{+0.72}_{-0.31}$ & 7.18$^{+1.16}_{-0.87}$ & 6.50 & 90/104\nl
N  720 & 0.58$^{+0.05}_{-0.06}$ & 0.10$^{+0.06}_{-0.04}$ & 2.16$^{+0.73}_{-0.72}$ & 1.42 & 64/56\nl
N 1052 & 2.88$^{+5.92}_{-1.25}$ & $<$1.09		 & 3.23$^{+5.57}_{-1.60}$ & 2.90 & 14/26\nl
N 1399 & 1.08$^{+0.02}_{-0.01}$ & 1.04$^{+0.26}_{-0.20}$ & 1.03$^{+0.25}_{-0.24}$ & 1.39 & 152/155\nl
N 1395 & 0.82$^{+0.04}_{-0.06}$ & 0.19$^{+0.13}_{-0.09}$ & 1.55$^{+0.85}_{-0.65}$ & 1.74 & 50/49\nl
N 1404 & 0.62$^{+0.03}_{-0.02}$ & 0.22$^{+0.06}_{-0.05}$ & 1.69$^{+0.34}_{-0.35}$ & 1.39 & 103/114\nl
N 1407 & 0.85$^{+0.07}_{-0.15}$ & 0.10$^{+0.07}_{-0.06}$ & 9.35$^{+7.73}_{-5.30}$ & 5.17 & 66/75\nl
N 1549 & 0.48$^{+0.26}_{-0.16}$ & 0.03$^{+0.11}_{-0.03}$ & 0.72$^{+1.40}_{-0.72}$ & 1.93 & 10/12\nl
N 2768 & 2.87$^{+16.13}_{-2.37}$& 0.20 			 & 1.87			  & 4.26 & 1.5/5\nl
N 3557 & 0.79$^{+0.12}_{-0.39}$ & $<$2.77		 & 8.04 		  & 8.04 & 5.0/5\nl
N 3585 & 0.40$^{+0.46}_{-0.13}$ & 0.11$^{+4.89}_{-0.11}$ & 5.38			  & 5.38 & 0.6/2\nl
N 3923 & 0.47$^{+0.16}_{-0.19}$ & 0.09$^{+0.10}_{-0.05}$ &17.90$^{+28.38}_{-2.77}$& 6.37 & 37/47\nl
N 4105\tablenotemark{c} & 0.76$^{+4.40}_{-0.69}$ & 0.2   & 6.10                   & 6.10 & 3.9/5\nl
N 4125 & 0.42$^{+0.14}_{-0.09}$	& 0.06$^{+0.05}_{-0.03}$ & 1.90 		  & 1.90 & 29/27\nl
N 4261 & 0.83$^{+0.05}_{-0.06}$ & 0.09$^{+0.06}_{-0.04}$ & 2.56$^{+0.71}_{-0.61}$ & 1.88 & 54/61\nl
N 4278 & 0.64$^{+0.33}_{-0.27}$ & 0.02$^{+0.24}_{-0.02}$ & 2.65$^{+4.20}_{-2.50}$ & 1.96 & 0.7/3\nl
N 4365 & 0.88$^{+0.60}_{-0.29}$ & 0.01$^{+0.13}_{-0.01}$ & 1.73$^{+1.38}_{-1.11}$ & 1.60 & 14/21\nl
N 4374 & 0.74$\pm$0.05          & 0.18$^{+0.16}_{-0.06}$ & 2.72$^{+0.69}_{-1.02}$ & 1.74 & 90/92\nl
N 4406 & 0.89$\pm$0.01  	& 0.37$\pm$0.05	 & 2.21$^{+0.24}_{-0.23}$ & 2.61 & 161/167\nl
N 4472 & 1.01$\pm$0.02  	& 1.12$^{+0.35}_{-0.24}$ & 1.17$^{+0.34}_{-0.32}$ & 1.64 & 166/160\nl
N 4486\tablenotemark{d} & 1.32$^{+0.03}_{-0.04}$ & 0.41$\pm$0.05          & 1.50$^{+0.65}_{-0.52}$ & 1.70 & 16/16\nl
N 4494 & 0.20$^{+0.08}_{-0.04}$ & 0.20  		 & 1.54 		  & 1.54 & 1.0/3\nl
N 4552 & 0.74$^{+0.07}_{-0.06}$ & 0.08$^{+0.07}_{-0.03}$ & 3.10$^{+0.77}_{-0.80}$ & 2.50 & 55/51\nl
N 4636 & 0.72$^{+0.03}_{-0.02}$ & 0.33$^{+0.15}_{-0.08}$ & 1.90$^{+0.49}_{-0.53}$ & 1.87 & 118/131\nl
N 4649 & 0.86$^{+0.02}_{-0.01}$ & 0.52$^{+0.32}_{-0.17}$ & 2.03$^{+0.78}_{-0.75}$ & 2.40 & 107/101\nl
N 4697 & 0.49$^{+0.12}_{-0.08}$ & $<$0.01		 & 2.78$^{+0.76}_{-0.77}$ & 2.54 & 62/65\nl
N 4696 & 1.32$^{+0.04}_{-0.05}$ & 0.86$^{+0.30}_{-0.22}$ & 10.15$^{+1.70}_{-1.24}$& 8.79 & 210/177\nl
N 5322 & 0.51$^{+0.18}_{-0.14}$ & 0.04$^{+0.08}_{-0.03}$ & $<$2.34		  & 1.87 & 20/21\nl
N 5846 & 0.76$^{+0.03}_{-0.04}$ & 0.29$^{+0.14}_{-0.09}$ & 4.89$^{+1.62}_{-1.15}$ & 4.23 & 157/148\nl
I 1459\tablenotemark{d}& 0.60$^{+0.12}_{-0.13}$ & 0.31$^{+0.66}_{-0.15}$ & 1.40			  & 1.40 & 13/24\nl
\enddata
\tablenotetext{a}{Errors are 90\% confidence}
\tablenotetext{b}{$N_H$ in units of 10$^{20}$ cm$^{-2}$}
\tablenotetext{c}{$Einstein$ IPC data used}
\tablenotetext{d}{A power law component was fit along with the Raymond-Smith plasma model}
\end{deluxetable}

\begin{deluxetable}{lllccccl}
%\small
\tablenum{2}
\tablewidth{0pt}
\tablecaption{Correlations}
\tablehead{
\colhead{} &
\colhead{Dependent} &  
\colhead{Independent} &  
\colhead{Slope} &  
\colhead{Dispersion}  & 
\colhead{N}  &  
\colhead{Prob\tablenotemark{a}} &
\colhead{Galaxies} \nl
\colhead{} &
\colhead{Variable} &  
\colhead{Variable} &  
\colhead{} &  
\colhead{(Dep. Var.)}  & 
\colhead{}  &  
\colhead{}  &
\colhead{Excluded}   
}
\startdata
1) & log $kT$    & log $\sigma$	& -0.20$\pm$0.36	& 0.26  & 30	& 0.034  \nl
2) & log $kT$    & log $\sigma$	& \phd1.45$\pm$0.20	& 0.10  & 27	& $9.5^{-5}$  & M32, N1052, N2768 \nl
\nl
3) & log $A$     & log $kT$	& \phd1.34$\pm$0.46	& 0.50  & 26	& $1.7^{-3}$ \nl
4) & log $A$     & log $kT$	& \phd2.44$\pm$0.69	& 0.48  & 24	& $3.1^{-3}$  & M32, N1052 \nl
\nl
5) & log $A$     & $Mg_2$ 	& -2.00$\pm$3.81	& 0.58	& 26	& 0.85	& \nl
6) & log $A$     & $Mg_2$ 	&  2.18$\pm$6.68	& 0.59	& 25	& 0.84	& M32 \nl
\enddata
\tablenotetext{a}{Probability that there is $no$ correlation, using Kendall's $\tau$ test;
dexponents are given as exponents.}
\end{deluxetable}

				\begin{figure}
				\title{
Figure Captions
				}
				\caption{
The $kT-\sigma$ correlation. The solid line is the fit to the data excluding 
the three galaxies with $kT>2$ keV (M32, NGC~1052, and NGC~2768). The dashed 
line shows the relation expected if the gas temperature and stellar kinetic 
temperatures were the same. The datum with the largest error bars (at log
$\sigma\approx 2.42$) is NGC~4105, the only galaxy in the plot where the 
temperature was determined using IPC data. 
				}

				\caption{
The abundance$-$temperature correlation.  The dashed line is the fit to the 
sample excluding the high temperature galaxies M32 and NGC~1052 (seen at right). 
Kendall's $\tau$ indicates the correlation is significant at the 99.84\% 
confidence level. 
				}

				\end{figure}

				
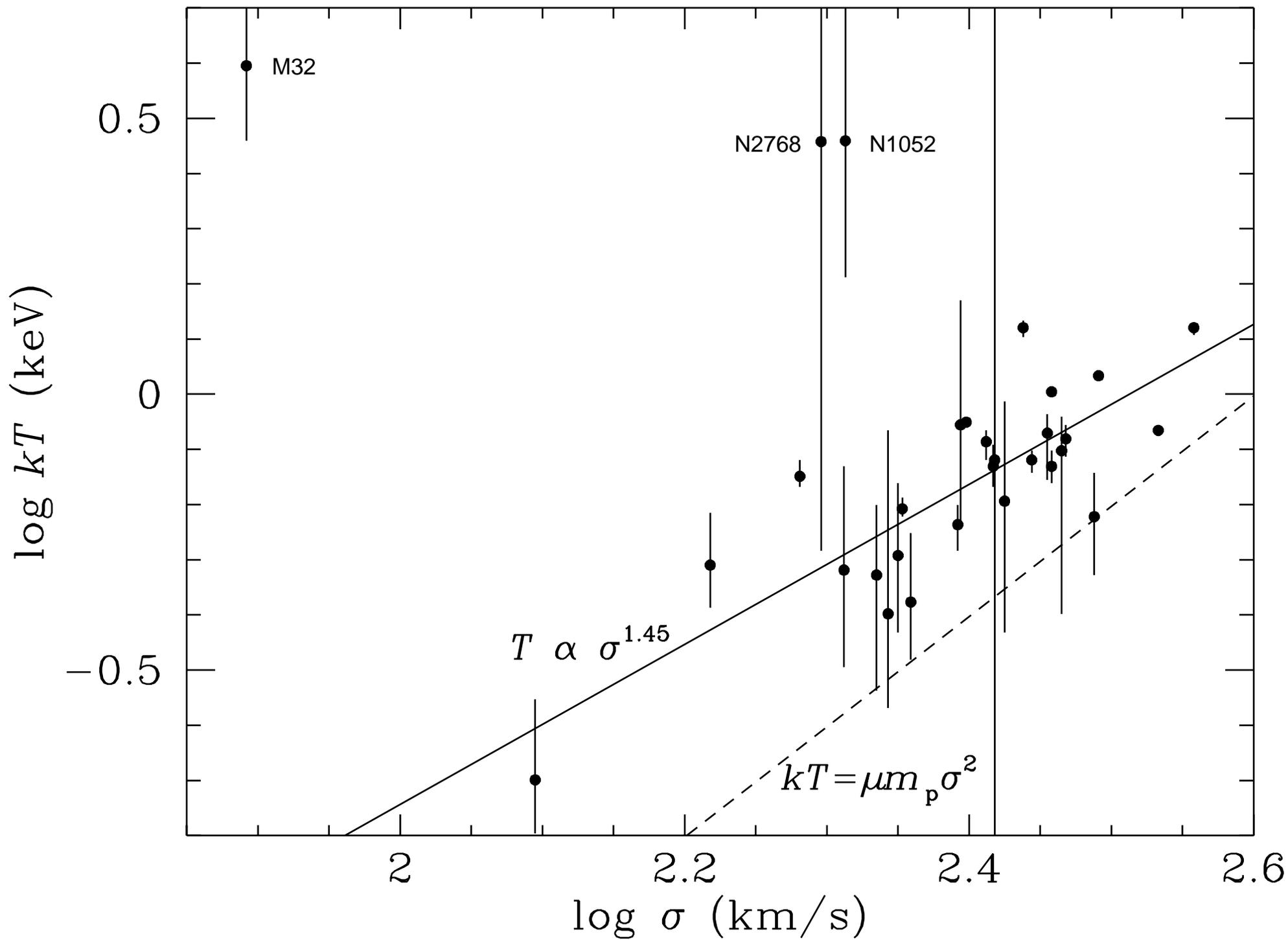
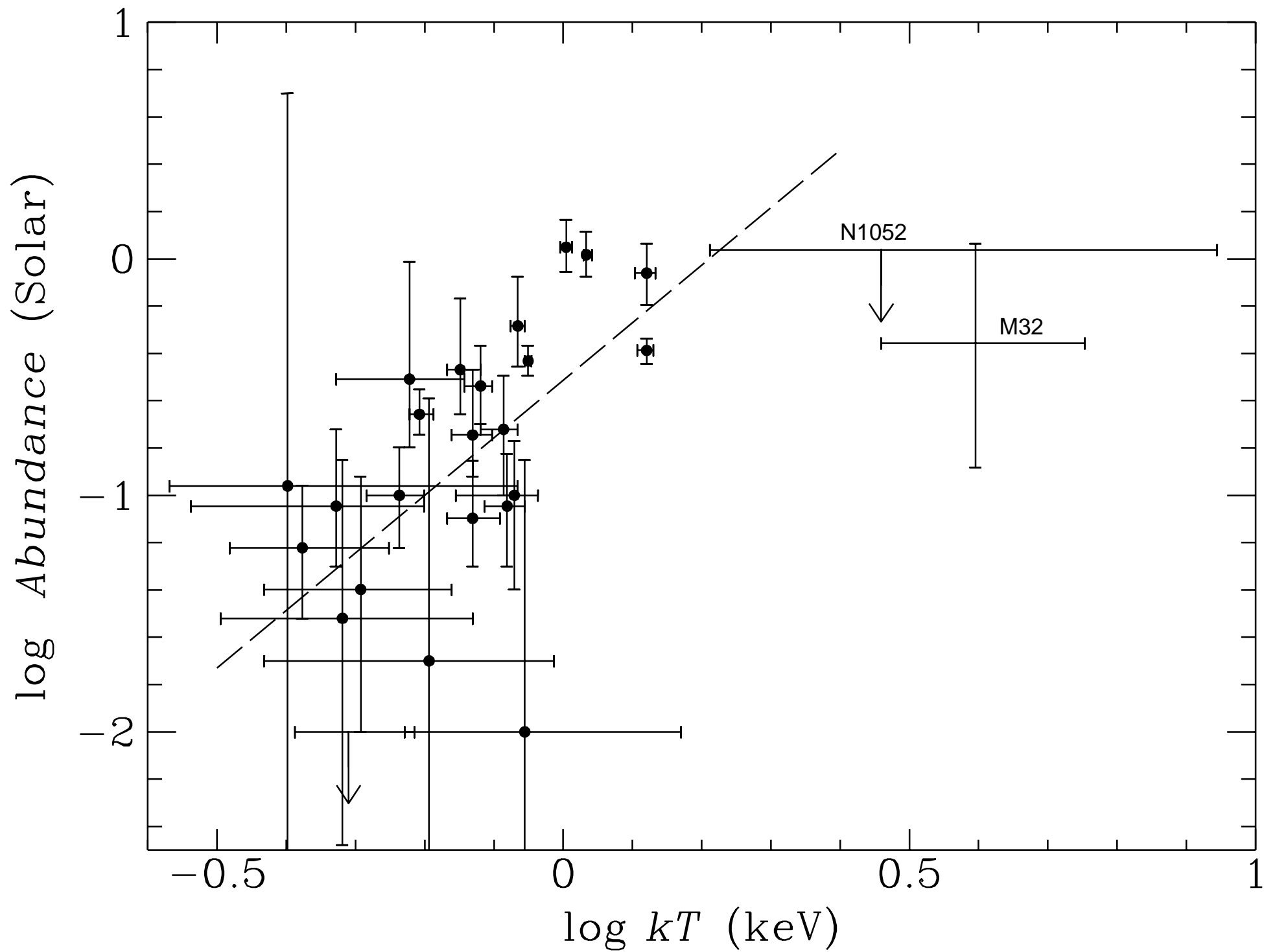
\end{document}